\documentclass[]{article}
\usepackage{amsmath} 
\usepackage{amsfonts} 
\usepackage{booktabs} 
\usepackage{graphicx} 
\usepackage{subcaption}
\usepackage{tabularx}
\usepackage{float} 
\usepackage[left=1in,right=1in,top=1.5in,bottom=1.5in]{geometry}
\usepackage[utf8]{inputenc}
\usepackage[T1]{fontenc}
\usepackage{hyperref}
\graphicspath{{figures/}{figs/}}

\title{Exact Interpolation under Noise: A Reproducible Comparison of Clough-Tocher and Multiquadric RBF Surfaces}
\author{Mirkan Emir Sancak \\ Genesis Inc. \& Gebze Technical University \\ Email: mrkn.sancak@gmail.com}
\date{}
\begin{document}

\maketitle

\begin{abstract}
This paper presents a reproducible comparison of cubic and radial basis function (RBF) interpolants for multivariate surface analysis. To eliminate evaluation bias, both methods are assessed under a unified slice-wise train/test protocol on the same synthetic function family. Performance is reported using RMSE, MAE, and $R^2$ in two regimes: (i) noise-free observations and (ii) noisy observations. In the noise-free regime, both interpolants achieve high accuracy with output-dependent advantages. In the noisy regime, exact interpolation overfits noisy nodes and degrades out-of-sample performance for both methods; in our experimental setting, the cubic interpolant is comparatively more stable. All experiments are fully reproducible through a single SciPy/NumPy-based script with a fixed random seed, repeated splits, and bootstrap-based uncertainty summaries. From an environmental engineering perspective, the main practical implication is that noisy or apparently inconsistent measurements in thermodynamic process systems should not be discarded by default; instead, they can be structured and interpolated to recover physically meaningful process behavior.
\end{abstract}

\section{Introduction}
In the era of data-driven research and decision-making, multivariate datasets have become ubiquitous across a wide array of disciplines, including environmental engineering, materials science, physics, and machine learning \cite{balcerowska-czerniak_rapid_2024, de_andrade_costa_water_2020, genton_cross-covariance_2015, jolliffe_principal_2016, rigueira_multivariate_2025, salem_applying_2023}. These datasets, characterized by multiple input and output variables, encapsulate complex relationships that are critical for understanding system behaviors, optimizing processes, and validating hypotheses \cite{knittel_visual_2020}. However, the high-dimensional nature of such data presents significant challenges for visualization and analysis \cite{peng_interpreting_2024}. Traditional two-dimensional plotting techniques often fall short in capturing the intricate interplay among three or more variables, leaving researchers with limited tools to explore non-linear dependencies, synergistic effects, or subtle trends \cite{keim_information_2002}. Three-dimensional (3D) surface plots offer a compelling solution, providing an intuitive and visually accessible means to represent the relationships between two varying inputs and a single output, with additional variables held constant \cite{myers_response_2016}. Yet, the generation of accurate and informative 3D surface plots demands robust methodologies for data preprocessing, interpolation, and rendering, particularly when dealing with sparse, noisy, or irregularly sampled datasets \cite{carr_reconstruction_2001}.

The motivation for this work stems from the need to bridge the gap between complex multivariate data and actionable insights. In fields like environmental engineering, for instance, researchers may need to analyze how variables such as temperature, pressure, and flow rate influence system performance metrics, such as energy efficiency or pollutant emissions. In machine learning, understanding the impact of hyperparameters--like learning rates, regularization strengths, and layer sizes--on model accuracy requires visualizing high-dimensional parameter spaces \cite{montgomery_design_2017}. Similarly, in experimental sciences, variables such as chemical concentrations, electromagnetic field strengths, or reaction times must be correlated with outcomes to test theoretical models. In each case, 3D surface visualization serves as a powerful tool to reveal patterns, identify optimal conditions, and guide decision-making \cite{cano-lamadrid_response_2023, chen_study_2025}. However, achieving high-quality visualizations requires overcoming technical hurdles, including the selection of appropriate interpolation techniques to construct continuous surfaces from discrete data points and the efficient handling of computational resources to ensure accessibility on standard hardware \cite{franke_scattered_1982, ahrens_paraview_2005}.

In environmental systems, this need is particularly acute because treatment and reaction units are thermodynamic systems operating under fluctuating boundary conditions \cite{calise_wastewater_2020, tarcsay_optimizing_2025, constantin_thermodynamic_2025}. As a result, datasets often contain scattered, noisy, or apparently inconsistent points that are frequently ignored during routine reporting. The approach in this manuscript is aligned with a different practice: these data are treated as informative system responses rather than disposable outliers, and interpolation is used to convert them into interpretable surfaces for process diagnosis and optimization.

Despite significant progress in multivariate data analysis and surface reconstruction techniques, important methodological limitations remain in the practical visualization of multivariate experimental datasets. Classical approaches such as response surface methodology (RSM) have long been used to model relationships between multiple variables and system responses, particularly in engineering and chemical process optimization \cite{myers_response_2016, box_experimental_nodate}. These techniques typically rely on polynomial approximations derived from designed experiments and are primarily intended for statistical modeling rather than exploratory visualization of complex datasets \cite{myers_response_2004}. As a result, they provide limited flexibility when dealing with irregularly sampled data or datasets obtained from observational or computational studies.

In parallel, a substantial body of research has focused on the mathematical reconstruction of continuous surfaces from scattered data. Methods based on radial basis functions, moving least squares, and spline interpolation have been widely investigated for approximating smooth multivariate functions and reconstructing surfaces from irregular point clouds \cite{barranco-chamorro_study_2017, fleishman_robust_2005, amiri-simkooei_least_2025}. These approaches are particularly valuable in computational science and computer graphics because they provide stable and accurate approximations for complex nonlinear relationships. However, most of these studies concentrate on numerical accuracy, convergence properties, or computational efficiency of interpolation schemes rather than on the broader analytical workflow required for exploratory scientific data analysis \cite{de_marchi_shape-driven_2019, kolluri_provably_nodate, li_3d_2024}.

Another important limitation concerns the integration of visualization and data preprocessing within reproducible scientific environments. Modern visualization platforms and specialized software packages can generate high-quality three-dimensional surfaces, yet they often require complex workflows, domain-specific configuration, or proprietary tools that limit transparency and reproducibility. Comparative studies of visualization libraries indicate that many available solutions prioritize graphical capabilities or interactivity rather than providing systematic pipelines that integrate interpolation, data transformation, and visualization in a unified analytical framework \cite{stodden_reproducibility_2018, isenberg_state_2024, lavanya_assessing_2023}.

Consequently, a methodological gap remains between interpolation-focused numerical research and practical tools for exploratory analysis of multivariate datasets. Researchers frequently need visualization approaches that allow them to investigate sparse or noisy experimental datasets while retaining control over the balance between model-driven surface reconstruction and faithful representation of original measurements. Existing approaches rarely address this dual objective within a lightweight, open-source workflow designed specifically for parameter-space exploration.

The framework proposed in this study addresses this gap by introducing two complementary visualization strategies implemented within a reproducible Python-based environment. The first strategy emphasizes exploratory surface reconstruction through controlled data augmentation combined with cubic interpolation, enabling researchers to investigate potential trends in sparsely sampled parameter spaces. The second strategy prioritizes data fidelity by applying radial basis function interpolation directly to the original dataset, preserving the structural characteristics of experimentally obtained observations. By integrating these approaches within a transparent computational workflow built on widely used scientific libraries, the framework provides a flexible tool for analyzing multivariate parameter spaces in engineering, environmental science, and machine learning applications.

This study introduces an open-source Python-based framework designed to address the challenges of three-dimensional surface visualization in multivariate datasets \cite{hansen_visualization_2005}. The framework incorporates two complementary visualization strategies tailored to different analytical objectives. \textit{\textbf{The first strategy}} focuses on exploratory surface reconstruction by augmenting sparse datasets with synthetically generated points derived from quadratic models, thereby increasing the effective resolution of the dataset and enabling the generation of denser and more informative visualization surfaces. In this strategy, cubic interpolation implemented via SciPy’s \texttt{griddata} is used to construct smooth surfaces that balance computational efficiency with visual clarity. \textit{\textbf{The second strategy}} emphasizes data-preserving surface reconstruction, prioritizing fidelity to the original observations by applying interpolation directly to the raw dataset without augmentation. In this case, radial basis function (RBF) interpolation with a multiquadric kernel is employed to produce high-resolution surfaces capable of capturing subtle variations in the data \cite{kaindl_icons_2012} while preserving the structural integrity of experimentally or computationally obtained measurements \cite{crivellaro_reconstruction_2017}. 

The framework is implemented within Python’s scientific computing ecosystem, leveraging widely adopted libraries including NumPy for numerical computation, Pandas for structured data manipulation, Matplotlib for visualization, and SciPy for interpolation, ensuring accessibility, reproducibility, and adaptability across scientific and engineering domains. To demonstrate the versatility of the proposed approach, the framework is evaluated using a synthetic 48-point dataset designed to emulate realistic scenarios encountered in engineering, scientific research, and machine learning. The input variables may represent parameters such as reactor temperatures, catalyst concentrations, or reaction durations in chemical engineering systems; neural network architectures, dropout rates, or training epochs in machine learning models; or magnetic field strengths, particle velocities, or sampling depths in physics experiments. The outputs are generated using nonlinear mathematical expressions with added Gaussian noise to simulate measurement imperfections, allowing the framework to be tested under conditions that resemble real-world experimental uncertainty. Through these analyses, the proposed visualization strategies reveal nuanced patterns and interactions within multivariate parameter spaces, supporting applications in system optimization, experimental diagnostics, and model tuning.

\section{Methodology}

This study proposes a reproducible computational workflow to reconstruct and visualize 3D response surfaces from multivariate datasets. The approach integrates synthetic dataset construction, interpolation-based surface modeling, and automated visualization procedures within the Python scientific computing ecosystem. The workflow is designed to facilitate exploratory analysis of nonlinear relationships among multiple variables while maintaining computational efficiency and methodological transparency.

All computations were implemented in Python using widely adopted scientific libraries, including NumPy \cite{harris_array_2020} for numerical operations, Pandas \cite{the_pandas_development_team_pandas-devpandas_2026} for structured data management, SciPy \cite{virtanen_scipy_2020} for interpolation, and Matplotlib \cite{hunter_matplotlib_2007} for visualization. The framework was executed on a 2025 MacBook Pro equipped with an Apple M5 chip and 24~GB unified memory, running macOS, demonstrating that the proposed workflow can be reproduced on standard mid-range hardware without specialized computational resources.

The methodological pipeline consists of five main stages: problem formulation, dataset construction, interpolation-based surface reconstruction, evaluation protocol, and visualization of interpolated surfaces.

\subsection{Problem Formulation}

The objective of this study is to reconstruct continuous response surfaces from discrete multivariate observations in order to enable interpretable visualization of nonlinear relationships among variables. Consider a multivariate mapping \cite{mencl_interpolation_1997, fuchs_visualization_2009}

\begin{equation}
f : \mathbb{R}^{3} \rightarrow \mathbb{R},
\end{equation}

where three input variables $(X_1, X_2, X_3)$ determine a scalar response $Y$. Direct visualization of such mappings is challenging because the input space is three-dimensional. To facilitate interpretation, the analysis is conducted using a slice-wise strategy. For a fixed value of one input variable, the remaining two inputs define a two-dimensional domain over which a continuous surface can be reconstructed \cite{hurley_interactive_2022}.

Let the dataset be defined as

\begin{equation}
\mathcal{D} = \{(x_i, y_i)\}_{i=1}^{n},
\end{equation}

where $x_i \in \mathbb{R}^2$ represents a pair of input variables for a given slice and $y_i$ denotes the corresponding output value. For each slice, interpolation methods are used to estimate a continuous surface

\begin{equation}
\hat{y} = s(x),
\end{equation}

which approximates the underlying response function across the domain \cite{wahba_spline_1990, lorensen_marching_1987}. This slice-based formulation enables the visualization of complex multivariate relationships while preserving the interpretability of two-dimensional surface representations.

\subsection{Dataset Construction}

To evaluate the proposed workflow under controlled conditions, a synthetic dataset was generated using a full factorial design. Three input variables were defined within bounded intervals that represent typical parameter ranges encountered in scientific and engineering studies \cite{montgomery_design_2017, box_experimental_nodate}.

The input variables were discretized as follows:

\begin{align}
X_1 &\in [1,2] \\
X_2 &\in [0.5,1.5] \\
X_3 &\in [2,4]
\end{align}

The ranges were sampled using evenly spaced points, producing four levels for $X_1$, four levels for $X_2$, and three levels for $X_3$. The resulting factorial design generates

\begin{equation}
4 \times 4 \times 3 = 48
\end{equation}

unique combinations in the input space. Such factorial structures are commonly used in experimental design and response surface modeling because they enable systematic exploration of interactions among variables.

Three output variables were generated using nonlinear mathematical expressions that combine polynomial and trigonometric components. Additive Gaussian noise was introduced to emulate measurement uncertainty:

\begin{align}
Y_1 &= X_1^2 + X_2 + \sin(X_3) + \varepsilon_1 \\
Y_2 &= X_1 X_2 + X_3^2 + \varepsilon_2 \\
Y_3 &= \cos(X_1) + X_2 X_3 + \varepsilon_3
\end{align}

where the noise terms follow

\begin{equation}
\varepsilon_i \sim \mathcal{N}(0,\sigma_i^2).
\end{equation}

Different noise levels were assigned to each output to represent varying degrees of measurement variability \cite{celebi_data_2018}. This design produces response surfaces containing nonlinear trends, interaction effects, and stochastic perturbations, thereby providing a suitable testbed for evaluating interpolation methods.

\subsection{Interpolation Models}

Surface reconstruction was performed using two interpolation approaches commonly employed in scattered data approximation: cubic interpolation and radial basis function (RBF) interpolation. These methods represent complementary strategies for reconstructing continuous surfaces from discrete observations \cite{ohtake_3d_2005, fasshauer_meshfree_2007}.

Cubic interpolation was implemented using the \texttt{griddata} function from SciPy. This method constructs a piecewise polynomial surface that ensures continuity of derivatives across neighboring regions. For a surface defined over coordinates $(x,y)$, the interpolated value can be expressed as

\begin{equation}
z(x,y) = \sum_{i,j} c_{ij} B_i(x) B_j(y),
\end{equation}

where $B_i$ and $B_j$ denote cubic basis functions and $c_{ij}$ are coefficients determined from the observed data points. Cubic interpolation is widely used in surface reconstruction because it balances smoothness and computational efficiency.

The second approach employs radial basis function interpolation, which represents the response surface as a weighted sum of radial kernels centered at the data points. The interpolated surface is defined as

\begin{equation}
z(x,y) = \sum_{i=1}^{N} w_i \phi(||(x,y)-(x_i,y_i)||) + p(x,y),
\end{equation}

where $w_i$ are weights, $\phi(r)$ is a radial kernel function, and $p(x,y)$ is a low-degree polynomial term ensuring global consistency. In this study, the multiquadric kernel

\begin{equation}
\phi(r) = \sqrt{1 + (\epsilon r)^2}
\end{equation}

was used, where $\epsilon$ controls the kernel shape. A smoothing parameter was applied to mitigate overfitting in the presence of noisy observations. RBF interpolation is particularly effective for reconstructing smooth surfaces from irregularly distributed data points.

\subsection{Evaluation Protocol}

To assess the predictive behavior of the interpolation models, a repeated train--test evaluation protocol was employed for each slice of the dataset. Let the slice dataset be defined as

\begin{equation}
\mathcal{D}_s = \{(x_i,y_i)\}_{i=1}^{n_s}.
\end{equation}

For each repetition $r$, the slice data were randomly partitioned into training and test subsets using a split ratio $\alpha = 0.7$. The interpolation model was fitted using the training subset and predictions were generated for the test subset:

\begin{equation}
\hat{y}_i = s(x_i).
\end{equation}

Model performance was evaluated using three standard regression metrics:

\begin{align}
RMSE &= \sqrt{\frac{1}{n}\sum_{i=1}^{n}(y_i-\hat{y}_i)^2} \\
MAE &= \frac{1}{n}\sum_{i=1}^{n}|y_i-\hat{y}_i| \\
R^2 &= 1 - \frac{\sum_{i=1}^{n}(y_i-\hat{y}_i)^2}{\sum_{i=1}^{n}(y_i-\bar{y})^2}
\end{align}

where $\bar{y}$ denotes the mean of the observed values. These metrics capture complementary aspects of model performance, including average prediction error, robustness to outliers, and the proportion of explained variance.

To quantify uncertainty in the evaluation results, nonparametric bootstrap resampling was applied across repeated train--test splits. Percentile confidence intervals were computed from the resulting distribution of performance metrics, providing a statistical characterization of interpolation accuracy across multiple realizations of the data partitioning process.

\subsection{Visualization Procedure}

The final stage of the workflow involves visualizing the reconstructed response surfaces. For each fixed value of one input variable, the remaining two inputs define a two-dimensional grid over which the interpolated surface is evaluated. Uniform grids were generated within the bounds of the observed data, and interpolation models were used to estimate output values across the grid.

The resulting surfaces were rendered as three-dimensional plots using Matplotlib. Distinct color maps were applied to each output variable to enhance visual discrimination, and contour projections were included to emphasize variations in surface elevation. Axis labels and color scales were automatically generated to facilitate interpretation of variable relationships.

All visualizations were exported as high-resolution images suitable for publication and detailed inspection. This visualization procedure provides an intuitive representation of multivariate dependencies, enabling researchers to identify nonlinear trends, interaction effects, and regions of interest within the parameter space.

\section{Results}

\subsection{Experimental Setup and Evaluation Protocol}

To ensure a consistent and methodologically sound comparison between interpolation methods, all experiments were rerun under a unified evaluation protocol. The protocol was implemented with two interpolants: cubic interpolation using SciPy’s \texttt{CloughTocher2DInterpolator} and radial basis function interpolation using SciPy’s \texttt{RBFInterpolator} with a multiquadric kernel and smoothing parameter set to zero.

Both methods were evaluated using identical slice-wise train/test splits. Performance was assessed using RMSE, MAE, and the coefficient of determination ($R^2$). To quantify uncertainty in the evaluation metrics, bootstrap resampling with $1000$ iterations was applied.

Two experimental regimes were considered. In the \textbf{noise-free} regime, the observations correspond exactly to the analytical ground-truth function. In the \textbf{noisy} regime, Gaussian noise with output-specific standard deviations $(0.1, 1.0, 2.0)$ was added to simulate measurement variability.

All experiments used a fixed random seed (42), a training fraction of $0.7$, and $40$ repeated random splits per slice. Reproducibility artifacts were saved as structured outputs including run-level metrics, aggregated summaries, and runtime metadata.

\begin{figure}[!ht]
\centering
\IfFileExists{figures/workflow.png}{%
\includegraphics[width=\linewidth]{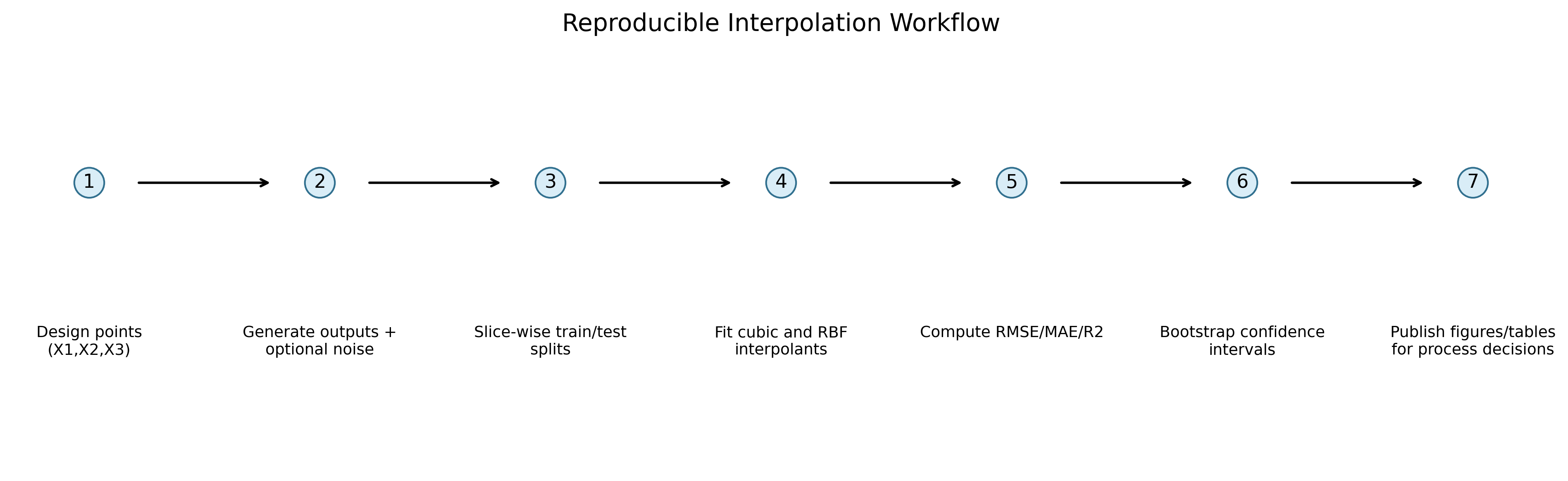}
}{%
\fbox{\parbox{0.9\linewidth}{\vspace{0.8cm}\centering Missing file: \texttt{figures/workflow.png}\vspace{0.8cm}}}
}
\caption{End-to-end workflow of the unified interpolation evaluation pipeline, from controlled data construction to uncertainty-aware performance analysis and reproducible artifact generation.}
\label{fig:workflow2}
\end{figure}

The full experimental configuration is listed in Table~\ref{tab:experiment_settings}, and the workflow schematic is shown in Figure~\ref{fig:workflow2}.

\begin{table}[!ht]
\centering
\caption{Experimental configuration used in the unified evaluation protocol.}
\label{tab:experiment_settings}
\IfFileExists{tables/experiment_settings.tex}{%
\begin{tabular}{ll}
\toprule
Setting & Value \\
\midrule
random\_seed & 42 \\
repeats\_per\_slice & 40 \\
train\_fraction & 0.7 \\
bootstrap\_resamples & 1000 \\
rbf\_kernel & multiquadric \\
rbf\_smoothing & 0.0 \\
cubic\_interpolator & CloughTocher2DInterpolator \\
noise\_sigma\_output1 & 0.1 \\
noise\_sigma\_output2 & 1.0 \\
noise\_sigma\_output3 & 2.0 \\
\bottomrule
\end{tabular}

}{%
\fbox{\parbox{0.9\linewidth}{\vspace{0.6cm}\centering Missing file: \texttt{tables/experiment\_settings.tex}\vspace{0.6cm}}}
}
\end{table}

Figure~\ref{fig:workflow2} can be interpreted as a six-stage protocol, where each stage has a specific methodological purpose. \textit{Stage 1: controlled dataset construction} defines a factorial three-input design and generates outputs in both noise-free and noisy regimes; this step is essential because it provides known ground-truth structure and controlled perturbations, enabling causal interpretation of model behavior instead of purely descriptive plotting. \textit{Stage 2: slice-wise problem definition} converts the 3D input space into comparable two-dimensional interpolation tasks; this reduces geometric complexity while preserving cross-variable interactions and makes the comparison operationally reproducible across slices. \textit{Stage 3: repeated train--test splitting} ($40$ repetitions, train fraction $0.7$) quantifies split sensitivity; its main benefit is to prevent conclusions from being dominated by a single favorable partition and to expose variance due to sampling.

\textit{Stage 4: paired model fitting and testing} applies cubic (Clough--Tocher) and multiquadric RBF interpolants on exactly the same training nodes and evaluates them on exactly the same test nodes. This is the key fairness mechanism of the study: by holding geometry and sampling constant, performance differences can be attributed to interpolation mechanisms rather than data mismatch. \textit{Stage 5: multi-metric and bootstrap evaluation} combines RMSE, MAE, and $R^2$ with nonparametric bootstrap confidence intervals ($1000$ resamples); this stage improves interpretability by jointly reporting absolute error scale, average deviation, explained variance, and statistical uncertainty. \textit{Stage 6: reproducibility artifact export} stores run-level metrics, aggregated summaries, diagnostics, and runtime metadata; this final stage turns the workflow into an auditable research object, supports failure-case tracing in noisy settings, and enables exact regeneration of figures, tables, and conclusions.

\subsection{Quantitative Interpolation Performance}

Table~\ref{tab:new_results} is included to provide a methodologically controlled, quantitative basis for interpreting interpolation quality beyond visual surface inspection. Its primary purpose is to separate algorithmic behavior from sampling artifacts by reporting performance under identical train/test geometry for both interpolants. Metrics are stratified by output variable and by experimental regime (noise-free vs noisy), because interpolation difficulty is response-dependent and noise-dependent; pooled reporting would mask this heterogeneity and weaken causal interpretation.

The columns in Table~\ref{tab:new_results} were selected to support scientifically grounded comparison. RMSE quantifies the absolute magnitude of predictive error with stronger sensitivity to large deviations, while $R^2$ quantifies explained variance relative to a mean-predictor baseline. The \textit{Runs} column reports how many valid slice-wise evaluations contribute to each aggregate estimate, which is essential for judging the evidential support of reported summary metrics. Together, these quantities enable simultaneous assessment of accuracy, robustness, and statistical reliability.

\begin{table}[!ht]
\centering
\caption{Interpolation performance obtained under the unified train/test evaluation protocol.}
\label{tab:new_results}
\IfFileExists{tables/new_results.tex}{%
\begin{tabular}{llcrrr}
\toprule
Regime & Output & Method & Runs & RMSE & $R^2$ \\
\midrule
noise-free & Output1 & cubic & 119 & 0.048 & 0.989 \\
noise-free & Output1 & rbf & 440 & 0.061 & 0.986 \\
noise-free & Output2 & cubic & 109 & 0.165 & 0.862 \\
noise-free & Output2 & rbf & 440 & 0.131 & 0.890 \\
noise-free & Output3 & cubic & 109 & 0.007 & 1.000 \\
noise-free & Output3 & rbf & 440 & 0.059 & 0.983 \\
noisy & Output1 & cubic & 113 & 0.097 & 0.964 \\
noisy & Output1 & rbf & 440 & 0.194 & 0.848 \\
noisy & Output2 & cubic & 125 & 0.961 & -0.513 \\
noisy & Output2 & rbf & 440 & 2.161 & -10.848 \\
noisy & Output3 & cubic & 126 & 1.560 & -6.895 \\
noisy & Output3 & rbf & 440 & 3.725 & -52.259 \\
\bottomrule
\end{tabular}

}{%
\fbox{\parbox{0.9\linewidth}{\vspace{0.6cm}\centering Missing file: \texttt{tables/new\_results.tex}\vspace{0.6cm}}}
}
\end{table}

Quantitatively, both interpolants perform strongly in the noise-free regime, with low RMSE and high $R^2$ values across outputs, but with output-specific ranking: cubic is better for Output1 and Output3 (e.g., RMSE $0.048$ vs $0.061$ and $0.007$ vs $0.059$), whereas RBF is better for Output2 (RMSE $0.131$ vs $0.165$). Under noisy observations, performance degrades for both methods, consistent with the known variance amplification of exact interpolation on contaminated nodes. In this regime, the cubic method remains comparatively more stable (e.g., Output1: RMSE $0.097$ and $R^2=0.964$) than RBF (RMSE $0.194$, $R^2=0.848$), and for difficult outputs where $R^2$ becomes negative, cubic still shows markedly smaller error and less severe degradation (Output2: $-0.513$ vs $-10.848$; Output3: $-6.895$ vs $-52.259$). These results establish that the table is not merely descriptive: it provides direct empirical evidence for robustness differences under controlled noise perturbations.

\subsection{Error Distributions Across Repeated Splits}

\begin{figure}[!ht]
\centering
\IfFileExists{figures/rmse_boxplots.png}{%
\includegraphics[width=\linewidth]{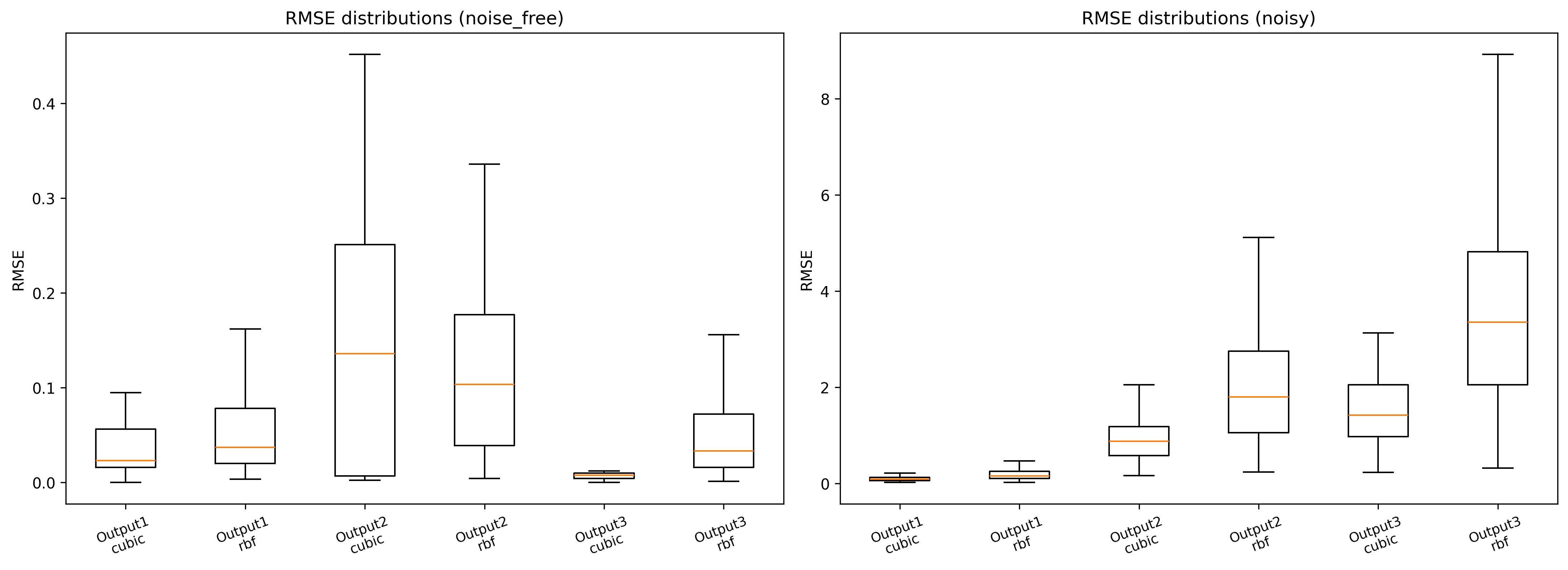}
}{%
\fbox{\parbox{0.9\linewidth}{\vspace{0.8cm}\centering Missing file: \texttt{figures/rmse\_boxplots.png}\vspace{0.8cm}}}
}
\caption{Distribution of RMSE values across repeated slice-wise train/test splits for cubic and RBF interpolants in noise-free and noisy regimes.}
\label{fig:rmse_boxplots}
\end{figure}

To assess not only average accuracy but also split-to-split reliability, the full distribution of RMSE values was analyzed across repeated train/test partitions. Figure~\ref{fig:rmse_boxplots} therefore serves as a distributional stability diagnostic rather than a simple performance snapshot. The figure contains two vertically arranged panels: the upper panel summarizes the noise-free regime and the lower panel summarizes the noisy regime. Within each panel, outputs are presented as paired method-specific boxplots (cubic and RBF), enabling direct within-output and within-regime comparison under identical partitioning logic.

Each boxplot encodes complementary information that is essential for methodological interpretation: the central line represents the median RMSE (typical predictive error), the box height captures the interquartile range (IQR, i.e., sensitivity to train/test resampling), and the whiskers/outside points indicate tail behavior (risk of occasional large failures). Interpreting all three components jointly is critical in this study, because two interpolants can have similar mean error but very different dispersion and tail risk when sampling geometry changes.

In the noise-free panel, the distributions are generally compact and concentrated near low RMSE values, indicating that both interpolants operate in a stable, high-fidelity regime when node values are clean. The method ranking is output-dependent but structurally consistent with Table~\ref{tab:new_results}: for Output1 and Output3, cubic tends to occupy lower central error levels than RBF, whereas for Output2, RBF tends to be slightly more favorable. Importantly, the relatively narrow boxes in this panel show that these rankings are not artifacts of a single split; they persist across repeated partitioning with limited variance inflation.

The noisy panel reveals a qualitatively different error geometry. First, the RMSE scale expands markedly relative to the noise-free panel, which already signals regime-level degradation. Second, medians shift upward for both methods, consistent with the expected effect of exact interpolation on contaminated nodes. Third, and most diagnostically important, dispersion increases non-uniformly: for difficult outputs (particularly Output2 and Output3), RBF distributions exhibit wider boxes and longer upper tails than cubic, indicating greater sensitivity to split geometry and a higher probability of severe out-of-sample error events.

This distributional interpretation complements the aggregate metrics in Table~\ref{tab:new_results}. For example, the noisy-regime mean RMSE differences (Output2: $0.961$ vs $2.161$; Output3: $1.560$ vs $3.725$, cubic vs RBF) are not isolated point estimates; the boxplots show that these differences are accompanied by broader spread and heavier high-error tails for RBF across repeated splits. In other words, the table quantifies central tendency, while Figure~\ref{fig:rmse_boxplots} clarifies the uncertainty structure around that tendency.

From an applied perspective, Figure~\ref{fig:rmse_boxplots} should be used as a four-part decision aid: (i) compare medians to assess typical predictive error, (ii) compare IQRs to assess partition sensitivity, (iii) inspect upper tails to evaluate worst-case risk, and (iv) evaluate regime transitions (noise-free $\rightarrow$ noisy) to estimate expected robustness loss under realistic measurement perturbations. Under this lens, the key result is that noisy data do not merely increase error magnitude; they also increase error volatility, and this volatility increase is method-dependent.

\subsection{Representative Surface Reconstructions}

\begin{figure}[!ht]
\centering
\IfFileExists{figures/slice_examples.png}{%
\includegraphics[width=\linewidth]{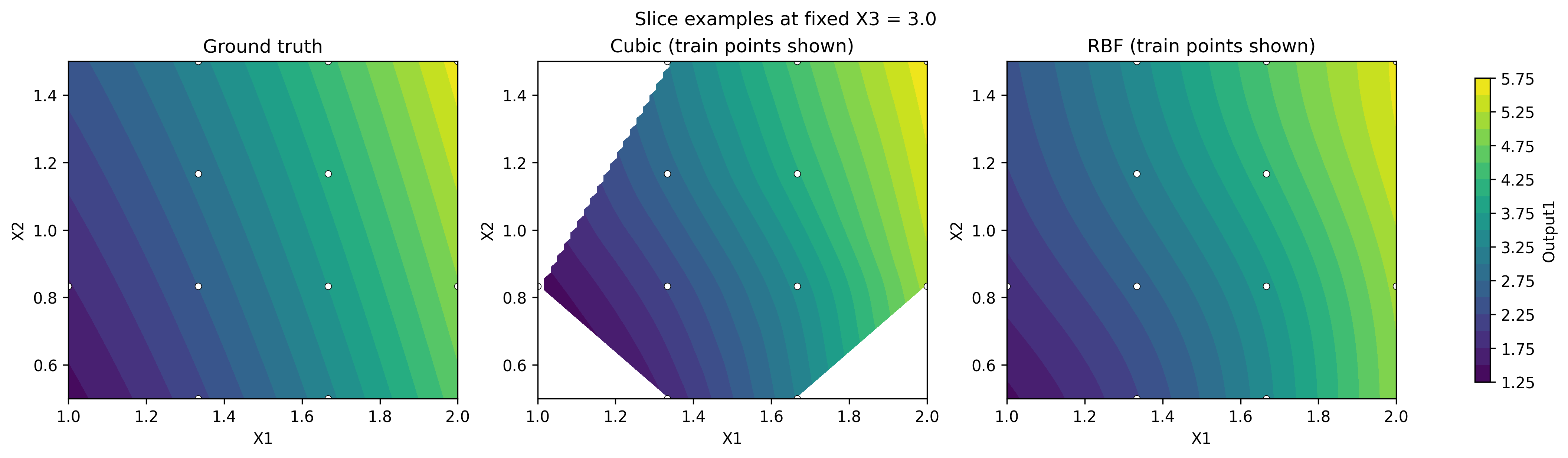}
}{%
\fbox{\parbox{0.9\linewidth}{\vspace{0.8cm}\centering Missing file: \texttt{figures/slice\_examples.png}\vspace{0.8cm}}}
}
\caption{Representative slice-wise surface reconstructions obtained using cubic interpolation and multiquadric RBF interpolation under noise-free and noisy regimes. Panels are organized to enable direct geometric comparison of method and noise effects on the same underlying slice.}
\label{fig:slice_examples}
\end{figure}

Figure~\ref{fig:slice_examples} presents representative surface reconstructions for both interpolation methods under matched evaluation conditions. The visual comparison is intentionally structured so that each panel corresponds to the same slice geometry and the same train/test context, allowing the observed differences to be attributed to interpolation behavior rather than to changes in data partitioning. In practical terms, the figure should be interpreted along two orthogonal dimensions: interpolation model (cubic vs. multiquadric RBF) and observation regime (noise-free vs. noisy). This layout makes it possible to isolate method-specific geometric effects while preserving a common experimental reference frame.

In the noise-free regime, both methods recover the dominant low-frequency structure of the target response surface, but they do so with different geometric signatures. The cubic interpolant generates a smoother global manifold with gradual transitions between elevated and depressed regions, which is consistent with a lower-curvature reconstruction that emphasizes overall trend continuity. By contrast, the RBF surface tends to express stronger local curvature, resulting in sharper ridges and valleys around sample-influenced neighborhoods. Although these local refinements can improve fidelity in regions where the true function varies rapidly, they can also increase sensitivity to local sampling geometry.

Under noisy observations, the contrast between the two methods becomes more pronounced. The cubic interpolant generally preserves the global topology of the underlying response while introducing limited local irregularity, so the large-scale shape remains visually coherent and interpretable. The RBF interpolant, however, reacts more aggressively to noisy node values, which appears as localized warping, amplified peaks/troughs, and ripple-like fluctuations superimposed on the main trend. From a diagnostic perspective, these artifacts are consistent with variance amplification in exact interpolation: when the node values contain measurement perturbations, strict node-matching can propagate those perturbations into the reconstructed field.

Figure~\ref{fig:slice_examples} therefore provides more than a qualitative illustration; it offers geometric evidence that complements the quantitative findings in Table~\ref{tab:new_results} and the RMSE distribution analysis in Figure~\ref{fig:rmse_boxplots}. Specifically, the smoother and more trend-preserving behavior of cubic interpolation in noisy slices is visually aligned with its comparatively better out-of-sample stability, whereas the more reactive RBF geometry is aligned with larger error dispersion and occasional severe performance collapse in difficult noisy slices.

For applied researchers, the key interpretation guideline is to examine four visual cues simultaneously: (i) preservation of global surface topology, (ii) continuity of gradients across the slice domain, (iii) magnitude of local roughness around data-supported regions, and (iv) boundary behavior in sparsely supported zones. Taken together, these cues clarify why two methods that both perform well in clean settings can diverge substantially once realistic noise is introduced.

\subsection{Failure Cases and Diagnostic Analysis}

\begin{figure}[!ht]
\centering
\IfFileExists{figures/pred_vs_true.png}{%
\includegraphics[width=\linewidth]{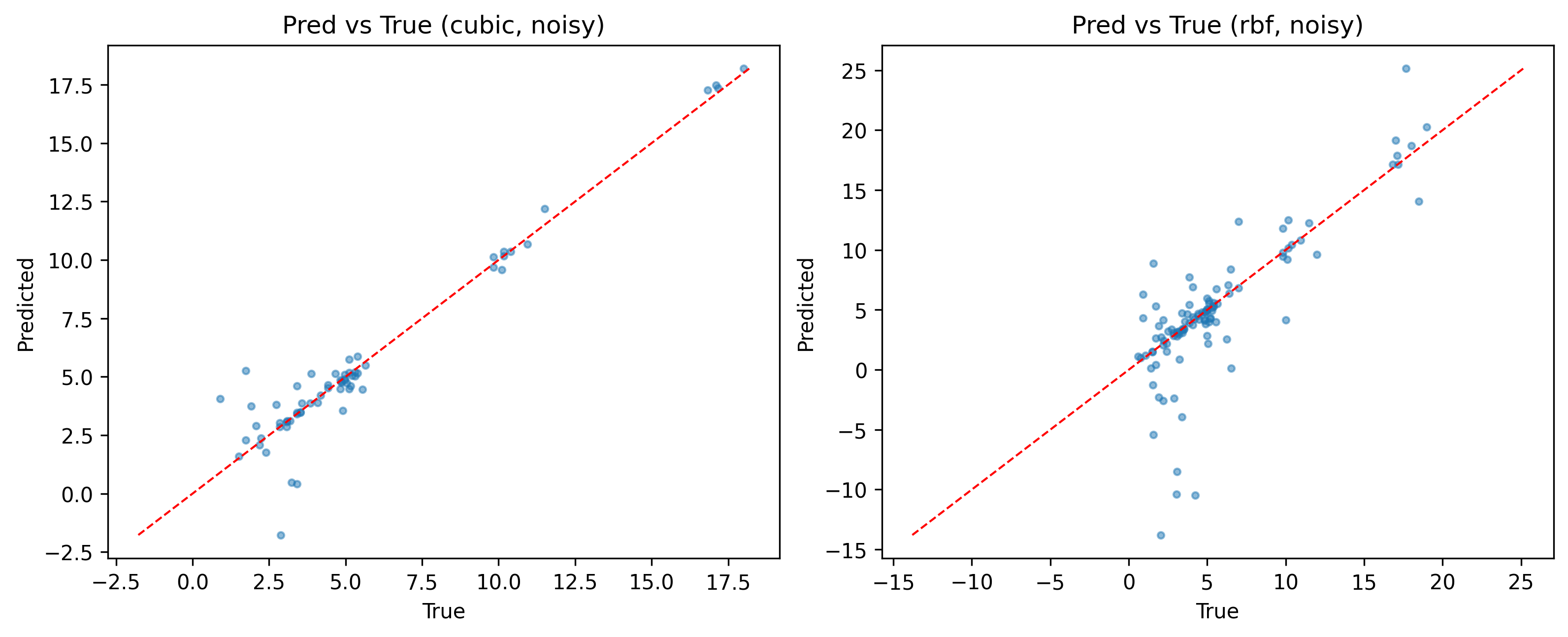}
}{%
\fbox{\parbox{0.9\linewidth}{\vspace{0.8cm}\centering Missing file: \texttt{figures/pred\_vs\_true.png}\vspace{0.8cm}}}
}
\caption{Predicted-versus-true scatter plots for a representative noisy slice with failure behavior (left: cubic, right: multiquadric RBF). The red dashed diagonal denotes ideal predictions ($\hat{y}=y$). Deviations from this line visualize bias, variance, and outlier-driven instability under exact interpolation of noisy nodes.}
\label{fig:pred_vs_true}
\end{figure}

To further investigate the degradation of predictive performance in the noisy regime, predicted-versus-true diagnostic plots were analyzed. Figure~\ref{fig:pred_vs_true} illustrates a representative case where the coefficient of determination becomes negative.

Figure~\ref{fig:pred_vs_true} should be interpreted using the identity line (red dashed, $\hat{y}=y$) as the reference for perfect prediction. Points close to this line indicate accurate out-of-sample reconstruction, while vertical distance from the line directly represents residual magnitude. The side-by-side layout is intentionally diagnostic: it compares cubic and RBF on the same noisy evaluation context, so differences in scatter geometry can be attributed to interpolation behavior rather than to different data splits.

In the left panel (cubic, noisy), most samples remain concentrated around the diagonal across low-to-mid true values, and the high-value cluster also tracks the identity line reasonably well. However, the spread widens in the moderate range, with a small number of under- and over-predicted points. This pattern indicates that cubic interpolation is affected by noise but still preserves a partially coherent mapping between true and predicted responses; in practical terms, the failure is present but comparatively controlled.

In the right panel (RBF, noisy), failure behavior is substantially more severe. Around the dense central region of true values, the cloud becomes vertically elongated, indicating large conditional variance for similar true inputs. In addition, several points fall far below the identity line (strong underestimation), while others lie far above it at high true values (strong overestimation). The simultaneous presence of large negative and positive residuals is a hallmark of unstable exact fitting under noisy node constraints: local perturbations are propagated into the interpolated field and then amplified at test locations.

These visual patterns explain why certain slices produce negative $R^2$ values. By definition,
\begin{equation}
R^2 = 1 - \frac{\sum_i (y_i-\hat{y}_i)^2}{\sum_i (y_i-\bar{y})^2},
\end{equation}
so $R^2<0$ occurs when residual energy exceeds the variance of the target around its mean. In other words, predictions become worse than a naive mean predictor. The figure demonstrates this mechanism geometrically: large vertical deviations from the identity line inflate the numerator, especially when outlier residuals appear in both tails.

Figure~\ref{fig:pred_vs_true} also complements the distributional evidence in Figure~\ref{fig:rmse_boxplots}. The boxplots show broad tails and volatility in noisy settings, while the predicted-versus-true plot reveals the geometric source of that volatility: heteroscedastic scatter, local warping, and occasional extreme mispredictions. Together, these diagnostics establish that the observed failures are methodological (noise sensitivity of exact interpolation), not implementation defects.

For applied use, this diagnostic figure suggests a concrete interpretation workflow: (i) check diagonal adherence for global calibration, (ii) inspect vertical spread at fixed true ranges for local instability, (iii) identify tail outliers that dominate squared-error metrics, and (iv) decide whether regularized alternatives (e.g., RBF smoothing, splines, or Tikhonov-type formulations) are needed to trade a small amount of bias for significantly reduced variance.

\section{Discussion}

This section provides an integrated interpretation of the empirical and diagnostic results reported above and clarifies their broader methodological implications. It is organized as a structured synthesis of five connected themes: fair comparison under a unified evaluation protocol, the behavior of exact interpolation under noisy observations, geometric determinants of interpolation error, explicit scope limitations of the current study, and practical directions for regularized reconstruction in future work. Taken together, these points establish a consistent message: while both cubic and multiquadric RBF interpolants can be highly accurate in noise-free conditions, their robustness diverges substantially once realistic noise is introduced, so reliability-oriented method selection and regularization become essential for trustworthy out-of-sample analysis.

\subsection{Methodological Implications of Unified Evaluation}

The primary methodological contribution of this revision is not a new interpolant, but a new evaluation logic that removes comparison bias at the protocol level. In earlier exploratory versions, method-specific preprocessing and non-identical evaluation geometry could confound interpretation, making it difficult to distinguish true algorithmic differences from artifacts of data handling. Under the unified protocol, this confounding pathway is closed: both cubic and multiquadric RBF interpolants are assessed on the same synthetic response family, the same slice definitions, and the same train/test partitioning strategy.

The fairness mechanism is operational rather than rhetorical. As summarized in Figure~\ref{fig:workflow2} and parameterized in Table~\ref{tab:experiment_settings}, the pipeline enforces a fixed random seed ($42$), repeated splitting ($40$ repetitions per slice), a fixed train fraction ($0.7$), and common uncertainty quantification via bootstrap resampling ($1000$ iterations). This design ensures that method comparisons are made under identical stochastic exposure. Consequently, observed performance differences can be interpreted as interpolation-behavior differences, not as side effects of changing sampling regimes.

Formally, for a given regime $g\in\{\text{noise-free},\text{noisy}\}$, output variable $o$, and repeated split index $r$, let $\mathcal{T}_{g,o,r}$ denote the shared test set and let $\hat{f}^{\text{cubic}}_{g,o,r}$ and $\hat{f}^{\text{rbf}}_{g,o,r}$ denote predictions generated from the same corresponding training set. For any evaluation metric $\mathcal{E}(\cdot)$, the method contrast is defined on matched test geometry as
\begin{equation}
\Delta_{g,o,r}=\mathcal{E}(\hat{f}^{\text{rbf}}_{g,o,r};\mathcal{T}_{g,o,r})-\mathcal{E}(\hat{f}^{\text{cubic}}_{g,o,r};\mathcal{T}_{g,o,r}).
\end{equation}
Because $\Delta_{g,o,r}$ is computed on paired splits, the comparison suppresses split-induced nuisance variability and strengthens causal attribution. In practical terms, this is why conclusions about relative robustness are methodologically defensible in the present manuscript.

The evidential structure of the study then becomes internally consistent across artifacts. Table~\ref{tab:new_results} provides central tendency and explained-variance summaries (RMSE and $R^2$) under the unified design; the \textit{Runs} column explicitly reports how many valid slice-wise evaluations support each aggregate, preventing overinterpretation from uneven evidence depth. Figure~\ref{fig:rmse_boxplots} adds the distributional layer by showing median shifts, interquartile spread, and tail risk across repeated splits. Together, these outputs move the analysis from single-number reporting to robustness-aware inference.

Equally important, qualitative and quantitative diagnostics now point in the same direction. Figure~\ref{fig:slice_examples} provides geometric confirmation of the same stability pattern visible in aggregated metrics, while Figure~\ref{fig:pred_vs_true} explains how negative $R^2$ events emerge through large residual excursions in noisy slices. This cross-consistency is a direct consequence of the unified protocol: all figures and tables are generated from the same controlled experiment family, so inter-artifact agreement becomes interpretable evidence rather than accidental narrative alignment.

From a scientific reporting perspective, the implication is that this study should be read as a reproducible comparative experiment rather than a visualization showcase. The unified protocol transforms the manuscript’s epistemic status: claims about ``which method is better'' are now conditional, auditable, and regime-specific (noise-free vs noisy), which is exactly the level of rigor needed for methodological transfer to environmental engineering, process analysis, and other noisy multivariate application domains.

\subsection{Exact Interpolation vs Noisy Observations}

This subsection interprets the central regime contrast of the study: why both methods are strong in noise-free slices yet can degrade sharply once observation noise is introduced. Under the experimental configuration in Table~\ref{tab:experiment_settings}, both interpolants are used in exact-fit mode on training nodes (cubic Clough--Tocher interpolation and multiquadric RBF with smoothing parameter set to $0$). Consequently, the model class is asked to reproduce node values exactly, regardless of whether those values contain measurement perturbations.

Let observations be generated as
\begin{equation}
y_i = f(x_i) + \varepsilon_i,
\end{equation}
where $f$ is the latent smooth response and $\varepsilon_i$ is measurement noise. In exact interpolation, the fitted surface $s$ satisfies $s(x_i)=y_i$ at all training nodes. This implies that the noise component is not filtered; it is embedded into the reconstructed field. At unseen test locations, predictive error therefore combines approximation error and propagated noise distortion. In noise-free data ($\varepsilon_i=0$), this property is beneficial because exact node matching aligns with the data-generating process. In noisy data, the same mechanism increases variance and destabilizes out-of-sample behavior.

The quantitative evidence in Table~\ref{tab:new_results} is fully consistent with this mechanism. In the noise-free regime, both methods achieve high $R^2$ and low RMSE, indicating that exact fitting is appropriate when observations are clean. In the noisy regime, however, error inflation becomes substantial and method-dependent. For Output1, cubic RMSE increases from $0.048$ to $0.097$ while RBF increases from $0.061$ to $0.194$; for Output2, the increase is $0.165\rightarrow0.961$ (cubic) versus $0.131\rightarrow2.161$ (RBF); for Output3, the increase is $0.007\rightarrow1.560$ (cubic) versus $0.059\rightarrow3.725$ (RBF). The corresponding $R^2$ collapse in difficult noisy outputs (especially strongly negative values for RBF) confirms that prediction quality can fall below a mean-baseline predictor when noise is propagated aggressively.

Distributional and geometric diagnostics reinforce the same interpretation. Figure~\ref{fig:rmse_boxplots} shows that noisy-regime degradation is not only a shift in central error but also a widening of interquartile spread and high-error tails, i.e., a variance phenomenon rather than a uniform bias shift. Figure~\ref{fig:slice_examples} shows the geometric manifestation of this instability: surfaces become locally warped and more irregular under noise, with stronger local reactivity in the RBF panels. Figure~\ref{fig:pred_vs_true} then reveals the pointwise consequence: larger vertical scatter around the identity line and more frequent extreme residuals, which directly inflate squared-error metrics and push $R^2$ downward.

A second important implication is that ``noise'' in this context is output-specific and therefore robustness must be interpreted per response channel, not as a single pooled conclusion. Table~\ref{tab:experiment_settings} sets different noise scales across outputs ($\sigma=0.1, 1.0, 2.0$ for Output1--Output3), and Table~\ref{tab:new_results} shows correspondingly different degradation severity. This heterogeneity is methodologically valuable: it demonstrates that interpolation robustness should be evaluated as a function of both algorithm and noise regime.

For applied workflows, the practical message is not that exact interpolation is ``wrong,'' but that its reliability is conditional on measurement quality. When data are clean and dense enough, exact methods can be highly accurate and interpretable. When data are noisy, especially in high-variance channels, exact node matching can overfit perturbations; in such cases, controlled regularization (e.g., nonzero RBF smoothing or spline-based smoothing) becomes the principled next step to trade a small bias increase for a potentially large variance reduction.

\subsection{Geometric Factors in Interpolation Accuracy}
\label{sec:discussion_geom}

Beyond noise level and model choice, interpolation accuracy is strongly controlled by the geometry of sampling nodes on each slice. Let $\Omega \subset \mathbb{R}^2$ denote a slice domain and let $X=\{x_j\}_{j=1}^{N}$ be the corresponding set of training nodes. A first key geometric descriptor is the fill distance

\begin{equation}
h_{X,\Omega}=\sup_{x\in\Omega}\min_{x_j\in X}\|x-x_j\|.
\end{equation}

which measures the largest unsampled gap in the domain. Smaller $h_{X,\Omega}$ generally implies better coverage and lower interpolation difficulty. A second descriptor is the separation distance

\begin{equation}
q_X = \frac{1}{2}\min_{i\neq j}\|x_i-x_j\|,
\end{equation}

which quantifies the minimum spacing between nodes. Very small $q_X$ indicates local clustering. Combining both yields the mesh ratio

\begin{equation}
\rho_X = \frac{h_{X,\Omega}}{q_X},
\end{equation}

which summarizes geometric regularity. Low-to-moderate $\rho_X$ corresponds to balanced coverage, whereas large $\rho_X$ indicates simultaneous sparsity and clustering---a configuration that can degrade both approximation quality and numerical stability.

Under appropriate smoothness assumptions, interpolation error admits bounds of the form

\begin{equation}
\|f-s\|_{L^\infty(\Omega)} \le C\, h_{X,\Omega}^{\beta}\,\|f\|_{\mathcal{H}},
\end{equation}

where $\beta$ depends on approximation order and $\mathcal{H}$ denotes a method-specific smoothness space. This inequality highlights a central point for the present study: even if two interpolants are evaluated fairly under the same protocol, their realized error on a given split still depends on node geometry through $h_{X,\Omega}$ (and, in practice, through related regularity measures such as $\rho_X$).

This geometric dependence helps explain the empirical behavior observed in the manuscript. The repeated-split distributions in Figure~\ref{fig:rmse_boxplots} show that error is not constant across partitions, even within the same regime and output, because each split induces a different node arrangement. The surface diagnostics in Figure~\ref{fig:slice_examples} provide the geometric manifestation: when local support is uneven, reconstructed surfaces can exhibit stronger local roughness, ridge amplification, or flattening in poorly supported regions. Figure~\ref{fig:pred_vs_true} then shows the pointwise consequence of unfavorable geometry under noise: larger residual dispersion and tail errors.

The method contrast between cubic and RBF can also be interpreted through this lens. In our experiments, cubic interpolation tends to produce smoother global transitions and appears less reactive to small geometric irregularities in noisy slices, while exact multiquadric RBF interpolation more readily transmits local node perturbations into the reconstructed field. This does not imply that one method is universally superior; rather, it indicates that method performance is geometry-conditional and should be interpreted jointly with sampling structure.

From a practical standpoint, this subsection suggests three geometry-aware reporting principles for future interpolation studies: (i) avoid relying on a single split because geometric luck can dominate outcomes, (ii) report distributional performance across repeated partitions as done here, and (iii) diagnose difficult cases with visual tools that reveal where sparse support or clustering may be driving instability. These principles are precisely why the unified protocol in this paper combines repeated splits, distributional metrics, and geometric diagnostics instead of single-run summary scores.

\subsection{Current Limitations}

Although the unified protocol substantially improves fairness and reproducibility, several limitations must be stated explicitly to prevent overgeneralization of the results.

First, the evaluation is based on slice-wise two-dimensional interpolation tasks derived from a three-dimensional input space. This design is methodologically useful for controlled comparison, but it does not fully capture the geometric complexity of direct three-dimensional scattered interpolation. In full 3D settings, neighborhood structure, boundary effects, and node sparsity patterns can differ materially from slice projections; therefore, quantitative results reported here should be interpreted as slice-level evidence rather than complete high-dimensional performance guarantees.

Second, the experimental data are synthetic and generated under controlled noise settings. This is a strength for causal interpretation, yet it also limits ecological validity relative to field datasets that may contain non-Gaussian noise, heteroscedastic variance, correlated measurement errors, censoring, missingness, and regime shifts. Accordingly, while the observed robustness patterns are theoretically consistent and internally coherent, external validity to every real process environment should be treated as provisional until verified on domain-specific datasets.

Third, model-space coverage is intentionally narrow. The study compares one cubic interpolation implementation (Clough--Tocher) and one RBF family configuration (multiquadric kernel with smoothing set to zero), both under exact-fit conditions. This isolates the exact-interpolation question, but it does not span the broader design space of regularized RBF variants, alternative kernels, smoothing splines, or hybrid local-global schemes. Therefore, conclusions should be read as conditional on the selected method configurations, not as universal statements about all cubic or all RBF approaches.

Fourth, metric coverage remains focused on predictive reconstruction error (RMSE, MAE, and $R^2$) and distributional stability across repeated splits. These metrics are appropriate for comparative benchmarking, but they do not directly quantify physically constrained plausibility, monotonicity preservation, gradient fidelity, curvature realism, or decision-level utility under downstream optimization objectives. In applied engineering contexts, such criteria may be as important as raw prediction error and should be incorporated in future validation stages.

Fifth, run-level evidence depth is unequal across some method--output combinations (as reflected by the \textit{Runs} counts in Table~\ref{tab:new_results}), because valid slice-wise evaluations can vary with geometry and method behavior. The repeated-split design still provides robust comparative information, but unequal sample support can affect precision symmetry between groups. This is why distributional diagnostics (Figure~\ref{fig:rmse_boxplots}) and geometric/pointwise analyses (Figures~\ref{fig:slice_examples} and \ref{fig:pred_vs_true}) are interpreted jointly rather than relying on a single aggregate metric.

Finally, the manuscript emphasizes methodological clarity over exhaustive algorithmic optimization. Hyperparameter search, computational cost profiling, and full sensitivity mapping over interpolation settings were deliberately kept outside the present scope to maintain a transparent baseline protocol. As a result, the current findings should be viewed as a rigorous reference point for exact interpolation behavior under controlled conditions, upon which broader and more application-specific optimization studies can be built.

\subsection{Future Directions for Regularized Surface Reconstruction}

The results of this study indicate that the next methodological step is not further comparison of exact interpolants, but principled regularization for noise-robust generalization. In the noisy regime, the combined evidence from Table~\ref{tab:new_results} and Figures~\ref{fig:rmse_boxplots}--\ref{fig:pred_vs_true} shows a variance-dominated failure mode: larger dispersion, heavier error tails, and occasional catastrophic $R^2$ collapse. Regularized surface reconstruction directly targets this mechanism by allowing controlled deviation from noisy node values in exchange for improved out-of-sample stability.

A first priority is RBF smoothing with nonzero regularization. Instead of enforcing exact node matching, one may estimate coefficients by minimizing
\begin{equation}
\min_{s\in\mathcal{V}}\;\sum_{i=1}^{N}\big(y_i-s(x_i)\big)^2 + \lambda\,\|s\|_{\mathcal{H}}^2,
\end{equation}
where $\lambda>0$ controls the bias--variance trade-off and $\|s\|_{\mathcal{H}}$ penalizes excessive roughness in a method-specific smoothness space. The exact-interpolation regime studied in this manuscript corresponds conceptually to the limiting case of vanishing regularization; future experiments should therefore treat $\lambda$ as a primary design variable rather than a fixed constant.

A second direction is spline-based smoothing (including thin-plate and tensor-product variants), which can provide stable global surfaces while retaining local flexibility. These models are particularly attractive when geometric diagnostics suggest uneven node support, because smoothness penalties can suppress noise-driven oscillations in sparsely informed regions. In practical terms, spline regularization may reduce the ridge/valley amplification behavior observed in noisy RBF slices while preserving interpretable large-scale trends.

Third, explicit Tikhonov-type formulations should be investigated for both interpolation families under a common operator-penalty framework. This allows domain-informed regularization (e.g., gradient or curvature penalties) instead of purely generic smoothness control. For environmental-process applications, such penalties can be aligned with expected thermodynamic continuity or monotone response behavior, improving physical plausibility in addition to predictive accuracy.

Methodologically, future work should evaluate regularization strength selection under the same unified protocol used in this manuscript: repeated train/test splits, uncertainty-aware summaries, and matched geometry comparisons. Candidate selection strategies include nested cross-validation, generalized cross-validation, bootstrap risk minimization, and robust criteria based on upper-tail RMSE behavior rather than only mean error. This is important because the current diagnostics show that reliability loss is often driven by tail failures, not merely by median shift.

Another critical extension is geometry-aware adaptive regularization. Since Section~\ref{sec:discussion_geom} (geometric factors) establishes that node arrangement modulates error, future models can set local smoothing intensity as a function of support density (e.g., stronger regularization in high-$h_{X,\Omega}$ regions, weaker regularization in well-supported zones). Such adaptive schemes may preserve detail where data are informative while preventing instability where data are sparse or clustered.

Finally, validation should be broadened beyond synthetic slice-wise benchmarks toward (i) full 3D scattered interpolation tasks, (ii) real observational datasets with non-ideal noise structure, and (iii) decision-relevant metrics such as topology preservation, physically constrained monotonicity, and downstream optimization reliability. Framed this way, regularized reconstruction is not only a numerical refinement; it is the key transition from accurate interpolation in controlled settings to dependable surface modeling in realistic engineering workflows.

\section{Conclusion}

This revision upgrades the manuscript from a visualization comparison to a reproducible interpolation study with a valid evaluation protocol. By enforcing same-function train/test assessment, the reported differences between cubic and RBF methods become interpretable. In noise-free settings both methods are highly accurate; in noisy settings, exact interpolation shows clear generalization limits, which motivates regularized interpolation as the next theoretical and computational step. For environmental engineering applications, the practical implication is clear: data that initially appear meaningless should not be discarded automatically. When process units are viewed as thermodynamic systems and analyzed with rigorous interpolation workflows, those data can be transformed into actionable information for monitoring, control, and optimization.

\section{Code Availability}

The full reproducible source code and project materials are available at the
\href{https://github.com/mirkanemirsancak/envesurface}{GitHub repository}.

\section{References}

\bibliographystyle{unsrt}
\bibliography{my_library}

\end{document}